\documentclass[aps,prl,reprint,superscriptaddress,amsmath,amssymb,a4paper,twocolumn,longbibliography]{revtex4-1}

\usepackage[utf8]{inputenc}
\usepackage{amsmath}
\usepackage{amssymb,amsfonts,latexsym}
\usepackage{bm}
\usepackage[mathcal]{euscript}
\usepackage{graphicx}
\usepackage{epsfig}
\usepackage{color}
\usepackage{pifont,wasysym,marvosym}
\usepackage{textcomp}
\usepackage{comment}
\usepackage{epstopdf}
\usepackage{gensymb}
\usepackage{dcolumn}
\usepackage{hyperref}
\usepackage{physics}
\usepackage{lineno}

\begin{document}
\title{Non-reciprocal topological solitons in active metamaterials}

\author{Jonas Veenstra}
\affiliation{Institute of Physics, Universiteit van Amsterdam, Science Park 904, 1098 XH Amsterdam, The Netherlands}
\author{Oleksandr Gamayun}
\affiliation{London Institute for Mathematical Sciences, Royal Institution, 21 Albemarle St, London W1S 4BS, UK}
\author{Xiaofei Guo}
\author{Anahita Sarvi}
\author{Chris Ventura Meinersen}
\author{Corentin Coulais}
\affiliation{Institute of Physics, Universiteit van Amsterdam, Science Park 904, 1098 XH Amsterdam, The Netherlands}
  

\begin{abstract}
From protein motifs~\cite{PhysRevE.82.011916} to black holes~\cite{PhysRevD.107.084042}, topological solitons are pervasive nonlinear excitations that are robust and can be driven by external fields~\cite{dauxois2006physics}.
So far, existing driving mechanisms all accelerate
solitons and antisolitons in opposite directions~\cite{Bennett_JStatPhys1981,dauxois2006physics}. 
Here we introduce a local driving mechanism for solitons that accelerates both solitons and antisolitons in the same direction instead: non-reciprocal driving. To realize this mechanism, we construct an active mechanical metamaterial consisting of non-reciprocally coupled oscillators~\cite{brandenbourger2019non,Ghatak_PNAS2020,Chen_NatComm2021,Wang2022} subject to a bistable potential~\cite{Kochmann_Bertoldi_review,Nadkarni_PRE2014,Nadkarni_PRL2016,Nadkarni_PRB2016,Raney2016,janbaz2022slow}. 
We find that such nonlinearity coaxes non-reciprocal excitations---so-called non-Hermitian skin waves~\cite{Coulais_NatPhys2021, Bergholtz_RMP2021, Shankar_NatRevPhys2022, Hatano_PRL1996, Gong_PRX2018, MartinezAlvarez_PRB2018, Yao_PRL2018, brandenbourger2019non, Ghatak_PNAS2020, Chen_NatComm2021, Wang2022, Weidemann_Science2022}, which are typically unstable---into robust
oneway (anti)solitons. 
We harness such non-reciprocal topological solitons by constructing an active waveguide capable of transmitting and filtering unidirectional information.
Finally, we illustrate this mechanism in another class of metamaterials that displays the breaking of ``supersymmetry'' ~\cite{Chen2014,upadhyaya2020nuts} causing only antisolitons to be driven. Our observations and models demonstrate a subtle interplay between non-reciprocity and topological solitons, whereby solitons create their own driving force by locally straining the material.
Beyond the scope of our study, non-reciprocal solitons might provide an efficient driving mechanism for robotic locomotion~\cite{Brandenbourger_arXiv2021} and could emerge in other settings, e.g. quantum mechanics~\cite{Meier_NatureComm2016,Pucher_NatPhot}, optics~\cite{Pernet_NatPhys2022,DelPino_Nature2022,Wanjura_NatPhys2023} and soft matter~\cite{Zhao_NatPhys2023}.
\end{abstract} 

\maketitle

Non-reciprocal active matter consists of local, non-reciprocal and nonconservative interactions~\cite{Coulais_NatPhys2021,Bergholtz_RMP2021,Shankar_NatRevPhys2022}. It is described by odd, \emph{viz.} asymmetric, or non-Hermitian matrices and tensors. Such materials exist across a wide range of scales, from electron transport~\cite{Kunst_PRL2018,Gong_PRX2018,MartinezAlvarez_PRB2018,McDonald_PRX2018,Yao_PRL2018,McDonald_NatComm2020}, electronics~\cite{Ronny_NatPhys2020}, optomechanics~\cite{Mathew_Nat_Nanotech2020} and photonics~\cite{Zhong_NatPhys2020,Weidemann_Science2022} to  colloids~\cite{Bililign2021Motile}, driven emulsions~\cite{Poncet_PRL2022},  biophysics~\cite{TanNature}, mechanics~\cite{brandenbourger2019non,Rosa_NJP2020,Scheibner2020non,Chen_NatComm2021}, robotics~\cite{Brandenbourger_arXiv2021} and traffic~\cite{Nagatani_RepProgPhys2002}. 
The non-Hermitian skin effect is a striking wave phenomenon occuring in non-reciprocal active matter. It has been observed in quantum mechanics \cite{Hatano_PRL1996,Gong_PRX2018,MartinezAlvarez_PRB2018,Yao_PRL2018}, mechanics~\cite{brandenbourger2019non, Ghatak_PNAS2020,Chen_NatComm2021,Wang2022}, photonics~\cite{Weidemann_Science2022}, and optomechanics~\cite{DelPino_Nature2022,Wanjura_NatPhys2023} where waves are unidirectionally amplified and have a spectrum that is extremely sensitive to boundary conditions. 

\begin{figure*}[t!]
  \includegraphics[width=2\columnwidth,clip,trim=0cm 0cm 0cm 0cm]{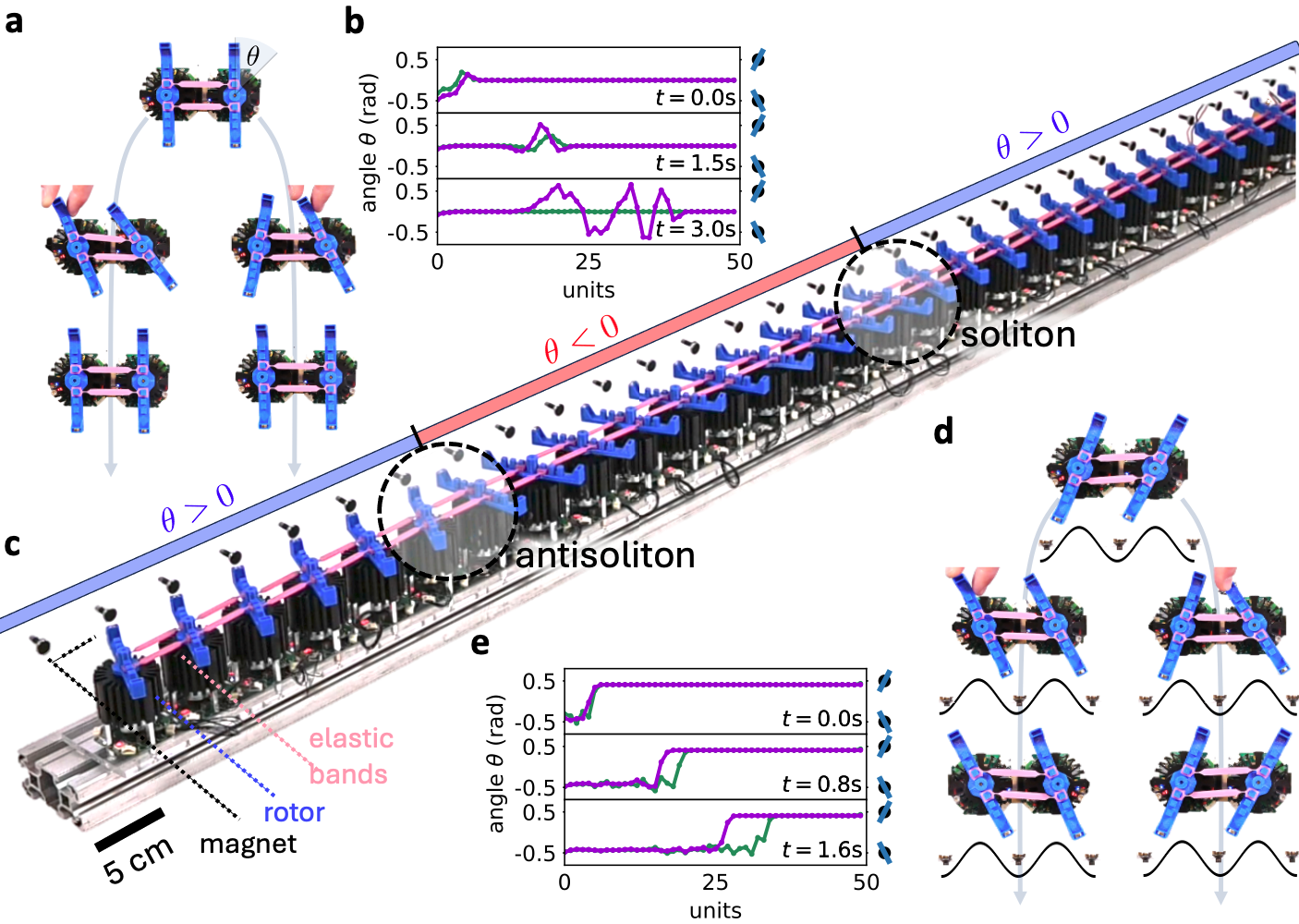}
 \caption{\textbf{Non-reciprocal topological solitons.}
 (\textbf{a}) Non-reciprocal response of a two coupled oscillators in the linear regime: actuation from the left induces a positive torque on the right while the actuation imposed from the right causes a negative torque on the left. The system returns to its equilibrium when the input displacement is removed.
 (\textbf{b}) When a chain of non-reciprocally coupled oscillators is perturbed, a wavepacket forms that is either unstable (purple) or dampened (green) depending on the balance between injected and dissipated energy.
 (\textbf{c}) The active metamaterial consisting of 50 elastically coupled motorized rotors---the picture shows half of the metamaterial for ease of visualization. When a magnet is added to the rotor tip and a periodic potential is generated by evenly spaced magnets, transition waves delimited by (anti)solitons can propagate along the chain.  
  (\textbf{d}) In the presence of periodically spaced magnets, each oscillator now has two stable configurations corresponding to the minima of the bistable potential. Switching the left configuration causes the right unit to follow suit while the same switch from the right does not propagate to the left.
 (\textbf{e}) When a soliton is seeded from the edge, it rapidly acquires a steady state velocity. The velocity increases with non-reciprocal coupling strength $\kappa^a$. Data shown in panels b and e correspond to $\kappa^a=1.6\cdot10^{-3} \mathrm{Nm \ rad}^{-1}$ (purple) and $\kappa^a=2.2\cdot10^{-3} \mathrm{Nm \ rad}^{-1}$ (green).}
 \label{fig1}
\end{figure*}

Most studies have however focused on the linear regime where non-Hermitian skin waves inexorably diverge or die out (Fig.~\ref{fig1}ab). A natural question is whether nonlinearities can be leveraged to stabilize wave phenomena in non-Hermitian systems~\cite{Coulais_NatPhys2021}. 
Mechanical metamaterials are a natural platform to adress this question. In particular, topological solitons in dissipative settings have been shown to be protected against damping and to robustly guide energy and information~\cite{Kochmann_Bertoldi_review,Nadkarni_PRE2014,Nadkarni_PRL2016,Nadkarni_PRB2016,Raney2016,KatiaNatComm,janbaz2022slow}. 
But so far they have only been studied under the effect of constant external driving, which immutably drives solitons and anti-solitons in opposite directions. 
The only exception is the case of passive stiffness gradient~\cite{Arrieta2018,KatiaNatComm} that pushes solitons and antisolitons in the same direction. Yet the lack of translation invariance causes a gradual loss of energy and limits scalability beyond a few unit cells (see Methods).

Here, we discover a subtle interplay between non-reciprocity and topological solitons that enables robust transmission of unidirectional signals. 
Topological solitons impose a local strain gradient. Coincidentally, non-reciprocity injects momentum proportionally to strain gradients. 
Therefore, non-reciprocal topological solitons induce their own driving force and push themselves in a direction that is independent of their topological charge.
Furthermore, we show that soliton and antisoliton velocities can be independently tuned by the nonlinearity of the metamaterial, which enables the material to exhibit robust waveguiding and filtering properties. Finally, we extend our findings to another type of metamaterial---the Kane-Lubensky chain~\cite{Chen2014,Ghatak_PNAS2020,upadhyaya2020nuts}---in which only antisolitons drive themselves. 
Our findings show how nonlinearities can be harnessed to promote topological excitations that stabilize the inertial dynamics of non-conservative systems \cite{Bililign2021Motile,TanNature,Braverman2021,Brandenbourger_arXiv2021}.

Our active mechanical metamaterial shown in Fig.~\ref{fig1}c consists of 50 3D printed rotating arms 
that are elastically coupled by rubber bands and positioned 
such that the $i^\mathrm{th}$ oscillator experiences a torque $\tau_i =\kappa (\theta_{i+1} + \theta_{i-1} -  2\theta_{i+1})$ (see Methods).
By coupling the torque on each oscillator antisymmetrically to the angle deviation of its neighbors according to $\tau_i^a = \kappa^a (\theta_{i-1}-\theta_{i+1})$ the system acquires a non-reciprocal 
response~\cite{brandenbourger2019non,Ghatak_PNAS2020,Wang2022}. Here $\tau_i^a$ denotes the active torque on the $i^{\mathrm{th}}$ oscillator,  $\kappa^a$ represents the non-reciprocal coupling strength  and $\theta_i$ is the angle deviation from the rest state.
In practice, this force rule means that actuating a pair of oscillators from the left causes an amplified response to the right, while the same actuation from the right causes the opposite response on the left (Fig.~\ref{fig1}a and Supplementary Video 1).
 

At the linear level, a finite oscillator chain exhibits non-Hermitian skin modes that amplify unidirectionally at all frequencies and exponentially localize towards the edge at a rate dependent on $\kappa^a$~\cite{MartinezAlvarez_PRB2018,Yao_PRL2018,LeePRB,McDonald_PRX2018}. In principle, this amplification imparts the metamaterial with 
an intrinsically unidirectional response.
Yet in practice its waveguiding capabilities are severely restricted
, since waves either blow up or die out unless non-reciprocity and
damping are meticulously tuned (Fig. \ref{fig1}b and Supplementary Video 1).

\begin{figure*}[t!]
 \begin{center}
  \includegraphics[width=2\columnwidth,height=1\columnwidth]{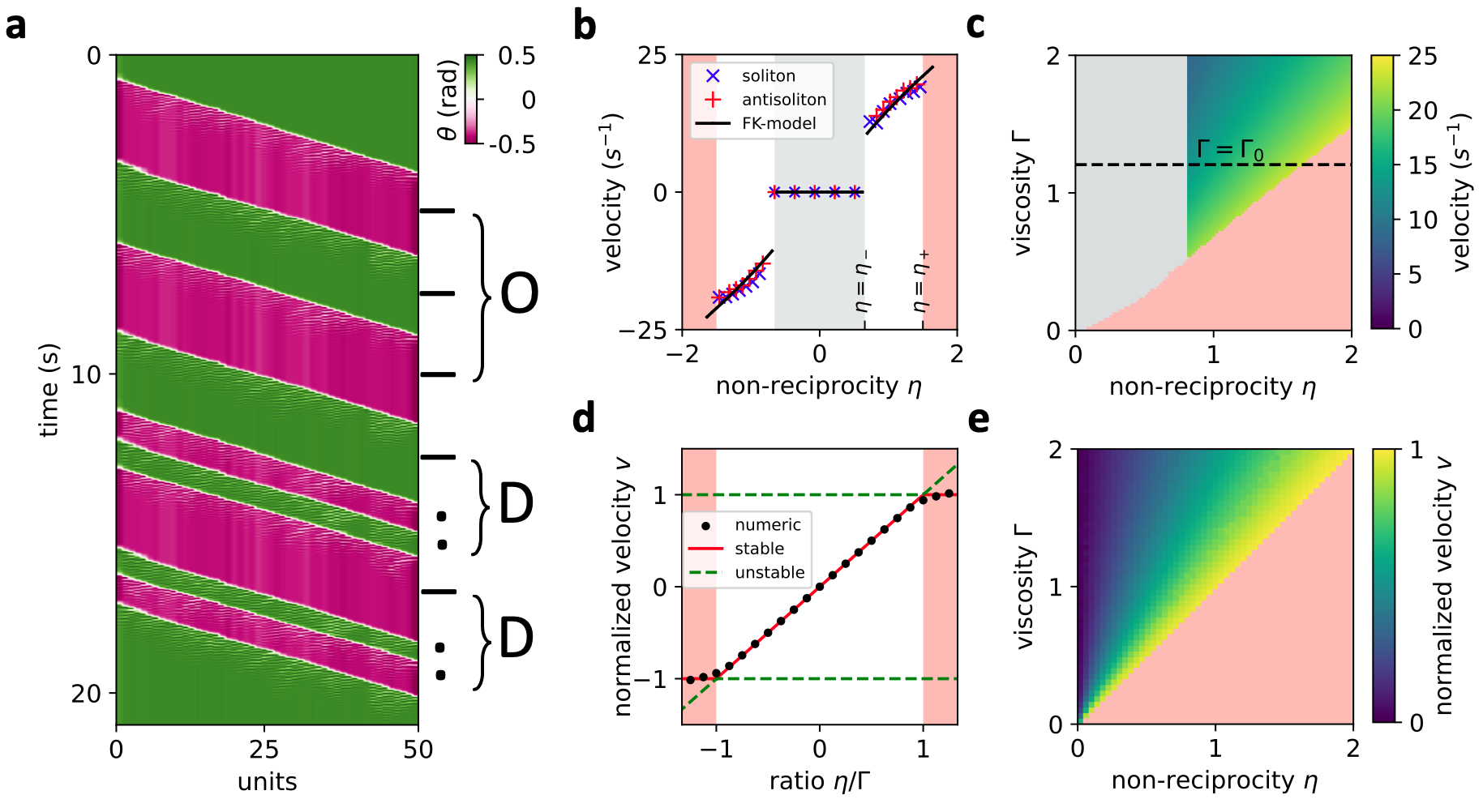}  
 \end{center}
 \caption{{\bf{Solitons and antisolitons travel in the same direction.}} (a) Experimental kymograph of soliton and antisoliton excitations from the edge propagating at equal and constant velocity along the material at intervals that encode the word `ODD' in Morse for $\kappa^a=1.6\cdot10^{-3} \mathrm{Nm \ rad}^{-1}$.
 (\textbf{b}) The (anti)soliton velocity observed experimentally and in simulation for a range of the non-dimensional activity $\eta$.  
 The solid line shows data from the non-reciprocal Frenkel-Kontorova model simulated with the experimental parameters $\Gamma = 1.3$ and $D=1.2$. The shaded areas denote regions bounded by thresholds $\eta_{\pm}$ where the metamaterial is unstable (light red) and where the (anti)soliton remains static (grey). The relative error of the velocity was smaller than $2\%$, found by averaging over $N=3$ runs for each datapoint.  
 (\textbf{c}) Phase diagram of the Frenkel-Kontorova model as a function of the rescaled non-reciprocity and viscous damping.
 The dashed line at $\Gamma_0 = 1.3$  corresponds to the viscous damping in the experiment. Notice that the jump in soliton velocity diminishes as $\Gamma$ increases, eventually vanshing in the overdamped limit.
 (\textbf{d}) Soliton velocity as a function of the ratio $\eta/\Gamma$ between activity and damping in the continuum model of Eq.~\eqref{eq:OddSineGordon} found by numerical integration. 
 The red and green lines show the steady state velocity as predicted by the stable and unstable fixed points of Eq.~\eqref{eq:NR_dynamical_equation} respectively. 
 (\textbf{e}) Phase diagram for solitons in the odd sine-Gordon equation, demonstrating velocity dependence on $\eta/\Gamma$. 
}
 \label{fig2}
\end{figure*}

To tame skin waves in the (strongly) nonlinear regime and turn them into topological solitons, 
we create a bistable potential by attaching magnets to the oscillator arms and to a periodic substrate (Fig. \ref{fig1}c). In this configuration, each oscillator now has two stable states where the magnetic, elastic and active torques balance instead of the single rest state in the linear case.  
When two bistable elements are coupled together and $\kappa^a$ is sufficiently large, switching stable states in one oscillator arm induces a transition in its neighboring oscillator while performing the reverse action does not bring about a switch. Crucially, owing to the bistable potential, the transition lasts even after the input displacement is removed (Fig.~\ref{fig1}d), contrary to the linear case.

When a switch is applied in an extended system of oscillators, a domino effect occurs that gives rise to a unidirectional transition wave with a distinctly soliton-like profile. The velocity of this travelling topological soliton, separating domains of left- and right-oriented oscillator arms, depends on $\kappa^a$ (Fig.~\ref{fig1}e). 
However, unlike toppling dominoes and 2-level systems with transition waves~\cite{Nadkarni_PRE2014,Raney2016,Deng2018,KatiaNatComm,janbaz2022slow,KatiaNatComm}, applying a reverse switch also induces a transition wave travelling at the same velocity, owing to the local injection of energy.
This behavior endows our metamaterials with robust unidirectional waveguiding capabilities, which we demonstrate by transmitting a message encoding the word ``ODD'' in Morse from one edge of the material to the other, without loss of amplitude or information (Fig.~\ref{fig2}a). This distinctive ability to continuously send trains of solitons and antisolitons provides an advantage over metamaterials based on constant driving~\cite{Raney2016,janbaz2022slow}, which have to be reinitialized by an antisoliton before a new soliton can be sent. \footnote{In addition, this data demonstates that non-reciprocal solitons and antisolitons can maintain their velocity over long distances. This property would be hard to achieve in the absence of non-reciprocal driving. Solitons and antisolitons could be sent in the same direction with carefully suited initial conditions, but they would irremediably slow down because of unavoidable dissipative effects.}

We experimentally investigate the response to solitons and antisolitons seeded at the edge of the chain for a range of the non-dimensional activity $\eta=2\kappa^a/\kappa \sqrt{D}$, where $D$ is the dimensionless amplitude of the bistable potential (see Methods) and find three regimes (Fig.~\ref{fig2}b).
Below a threshold at $|\eta|=\eta_-$, the active torque is not strong enough to overcome the hold of the magnetic potential and the soliton does not propagate into the material. For stronger non-reciprocity, excitations start to move spontaneously and acquire a velocity proportional to $\eta$, until a second threshold at $|\eta|=\eta_+$ is reached. At this point, (anti)solitons accelerate to the speed of sound (see Methods) and any further increase in the activity causes the excitations to become unstable and delocalize.

To rationalize our observations, we model the multistable active metamaterial with a non-reciprocal Frenkel-Kontorova chain: 
\begin{equation}
\Ddot{\phi}_i\!= \! \phi_{i-1}\!+\!\phi_{i+1}\!-\!2\phi_{i}\!-\!\frac{\eta}{2} (\phi_{i+1}\! -\! \phi_{i-1}) \! -\! \Gamma \dot{\phi_i} \!-\! D \sin(\phi_i) 
\label{eq:NRFK}
\end{equation}
Here, $\phi_i=2\pi\frac{\theta_i}{\theta_d} +\pi$ denotes the $i^\mathrm{th}$ oscillator angle normalized by the magnet spacing $\theta_d=1 \,\mathrm{rad}$ and shifted by $\pi$ while the nondimensional parameters $\eta$ and $\Gamma$ represent the non-reciprocity and dissipation (see Methods for details). For the range of amplitudes $-\pi<\phi_i<3\pi$ considered here, the force deriving from the bistable potential is well approximated by a sinusoidal function (see S.I. for details) with amplitude $D$~\footnote{In this range of amplitudes, we could equivalently model the nonlinear potential by a quartic potential. However, we will use later on the integrable nature of the sine-Gordon equation (the left hand side of Eq.~\ref{eq:OddSineGordon}) so opt for a sinusoisonal potential.}. The Frenkel-Kontorova model is known to host soliton solutions \cite{PEYRARD198488} that require a minimum energy to overcome the Peierls-Nabarro barrier in order to move along the lattice.
Models driven by a constant field have also been considered \cite{BraunPRE} where solitons and antisolitons move in opposite directions, contrary to the observations reported here.

We calibrate the experimental parameters with compression and oscillation experiments
(see Methods and S.I.) and find $\Gamma = 1.3\pm0.3$ and $D=1.2\pm0.3$. With these values, we numerically integrate Eq. \eqref{eq:NRFK} and find that it captures quantitatively and without free fit parameters the experimentally observed soliton velocity, the Peierls-Nabarro barrier and threshold of instability (Fig.~\ref{fig2}b).
A phase diagram reveals the ubiquity and tunability of unidirectionally travelling solitons (Fig.~\ref{fig2}c), confirming that the velocity generically increases with activity and decreases with dissipation. 

To get a better analytical understanding of the system, we now probe the continuum limit, where the lattice spacing is much smaller than the periodicity of the potential~\cite{PEYRARD198488}, the Peierls-Nabarro barrier decreases and eventually disappears (see Extended Data Fig. \ref{fig:PN_dependence}).
Here, the model of Eq.~\eqref{eq:NRFK} yields the sine-Gordon equation with an extra term that breaks spatial inversion symmetry and a dissipative term:
\begin{equation}
\pdv[2]{\phi}{t}-\pdv[2]{\phi}{x}+\sin \phi = -\eta\pdv{\phi}{x} - \Gamma\pdv{\phi}{t}.
\label{eq:OddSineGordon}
\end{equation}
Numerical integration of Eq.~\eqref{eq:OddSineGordon} confirms the linear dependence of the velocity on the 
ratio between non-reciprocity and
damping
for $|\eta| / \Gamma<1$ (Fig.~\ref{fig2}d). When this ratio exceeds $1$, Eq.~\eqref{eq:OddSineGordon} becomes unstable and high wavenumber radiative modes are amplified, although the wavefront velocity does not exceed the speed of sound (Fig.~\ref{fig2}e and see Methods for stability analysis).

By treating 
the nonreciprocal and damping
terms perturbatively and using the inverse scattering transform (see Methods), we analyze the time evolution of the (anti)soliton profile given by $\phi = \pm 4 \arctan \exp \frac{x-vt}{\sqrt(1-v^2)}$, known to be a solution to the standard sine-Gordon equation. We find a dynamical equation for the (anti)soliton velocity as a function of the ratio between non-reciprocity and
damping.
\begin{equation}
\frac{dv}{dt} = - (1-v^2)(\Gamma v - \eta) 
\label{eq:NR_dynamical_equation}
\end{equation}
Here, $v$ denotes the soliton velocity normalized by the speed of sound (see Methods). 
Eq.~\eqref{eq:NR_dynamical_equation} describes how solitons accelerate to a steady state velocity given by the stable fixed point $v=\eta/\Gamma$ below the threshold of instability. Beyond the threshold, there is a transcritical bifurcation where this fixed point becomes unstable. Another fixed point at the speed of sound $v=1$ then becomes stable, confirming numerical results (Fig.~\ref{fig2}d). 
In conclusion, 
the existence of non-reciprocal topological solitons is underpinned by stable fixed points, no matter how strong the non-reciprocal gain is. Notably, these topological solitons affect the physics of the non-Hermitian skin waves. They impose a gradient of strain, which in turn maintains a local non-reciprocal drive, no matter what the ratio between non-reciprocity and loss $\eta/\Gamma$ is.
This localized non-reciprocal driving confines non-Hermitian skin waves to the near vicinity of the soliton and hence nullifies the strong-sensitivity of the non-Hermitian skin effect to boundary conditions (See Extended Data Fig.~\ref{fig:PBC}).


\begin{figure*}[t!]
 \begin{center}
  \includegraphics[width=2\columnwidth]{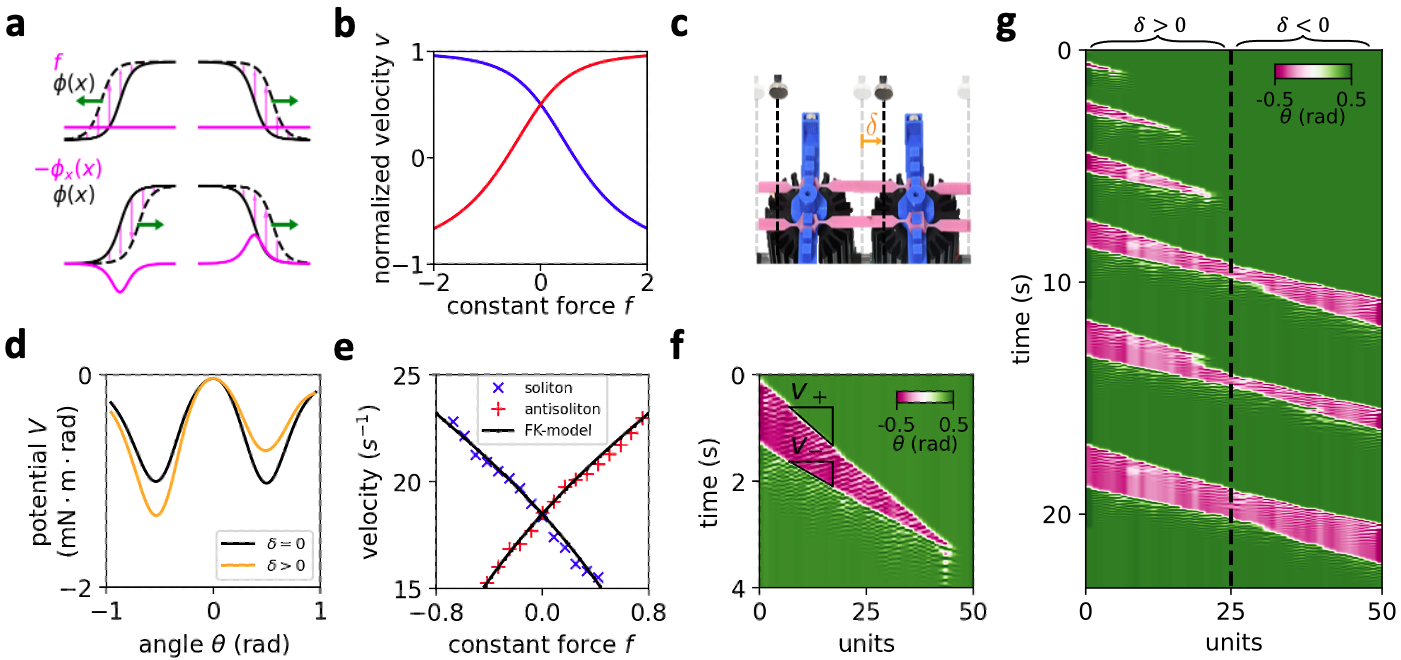}  
 \end{center}
 \caption{{\bf{Independent control of solitons and antisolitons.}}(\textbf{a}) A sketch comparing the effect of constant driving $f$ (top) and non-reciprocal driving $-\partial_x\phi$ (bottom) on soliton (left) and antisoliton (right) profiles in the sine-Gordon model. Black solid lines and dashed lines indicate the profile at times $t$ and $t+dt$ respectively. Magenta lines show the different driving fields, the magenta arrows indicate the discrete eigenmode of the soliton perturbation and green arrows show the resulting direction of propagation of the (anti)soliton.
 (\textbf{b}) Velocity based on the continuum prediction of Eq.~\eqref{eq:NRDC_steadystate_velocity} for solitons (blue) and antisolitons (red) vs. constant force $f$, for a non-reciprocity fixed at $\eta=0.5$. 
 (\textbf{cd}) A shift in the magnet position by an offset $\delta$ generates an asymmetric potential towards the left stable state. 
(\textbf{d}) Experimentally measured onsite potential vs. angle of the rotor for $\delta=0 \ \mathrm{mm}$ (black) and $\delta=4 \ \mathrm{mm}$ (orange).
(\textbf{e}) Experimental measurements of the terminal velocity of solitons (blue crosses) and antisolitons (red) vs. 
the constant force $f$. The black lines denote the numerical data given by the Frenkel-Kontorova model of Eq.~\ref{eq:NRFK} with an added constant force term (see Methods for details). The relative error of the velocity was smaller than $2\%$, found by averaging over $N=3$ runs for each datapoint.
(\textbf{f}) Solitons and antisolitons collide leading to annihilation for $f=0.4$. (\textbf{g}) Unidirectional nonlinear filter. Connecting two chains with opposite bias $\delta=\pm 3\mathrm{mm}$ together creates a low-pass filter for (anti)soliton excitations. 
The kymograph shows soliton and antisolitons excited at increasing time intervals. If the interval between soliton and antisolitons is smaller than some threshold, the signal annihilates before reaching the interface. At sufficiently large intervals, the signals are recovered at the  intervals on the other end of the chain. The data of (\textbf{efg}) was taken at a non-reciprocity of $\eta=1.1$.}
 \label{fig4}
\end{figure*}
At this point, we note that sine-Gordon solitons driven by a constant force $f$ have been studied extensively in the integrable systems literature~\cite{KivsharRevModPhys} and more recently in the mechanical metamaterials literature~\cite{Kochmann_Bertoldi_review,Nadkarni_PRE2014,Nadkarni_PRL2016,Nadkarni_PRB2016,Raney2016,janbaz2022slow}. 
Under constant driving, solitons and antisolitons move in opposite directions. For example, a positive constant driving pushes both solitons and antisolitons up, which drives the soliton backwards and the antisoliton forwards (Fig. \ref{fig4}a). In contrast, the non-reciprocal driving mechanism is the consequence of a subtle interplay between topological solitons and non-reciprocity.
On the one hand, topological (anti)solitons induce a local gradient of strain that is robust and whose sign is controlled by the topological charge of the soliton.
On the other hand, non-reciprocal driving injects momentum proportionally to the gradient of strain. Therefore, solitons locally induce their own driving force, of the form $\partial_x\phi\sim \sech x$, which precisely matches the discrete eigenmode of the spectrum of linear perturbations to the soliton profile~\cite{Kivshar1998}. Hence when $\eta>0$, this driving leads to an effective force that pushes the soliton (antisoliton) down (up). In turn, these two opposite forcings drive both solitons and antisolitons forward even though they have opposite topological charges.


Strikingly, combining both drives grants control over soliton and antisoliton velocities individually.
Repeating the inverse scattering transform on Eq.~\eqref{eq:OddSineGordon} plus a constant $f$ (see Methods) adds an extra term to Eq.~\eqref{eq:NR_dynamical_equation}:
\begin{equation}
	v_{\pm} = \frac{\pm \frac{\pi f}{4}\sqrt{\Gamma^2 - \eta^2 + \pi^2 f^2/16}+ \eta \Gamma}{\Gamma^2 + \pi^2 f^2/16}.
\label{eq:NRDC_steadystate_velocity}
\end{equation}
Here, $v_+$ and $v_-$ denote the soliton and antisoliton velocities respectively, which depart from one another as the constant driving $f$ is increased (Fig.~\ref{fig4}b). 
Experimentally, we realize this by biasing the periodically spaced magnets with respect to the oscillators (Fig.~\ref{fig4}c) by an offset $\delta$. This introduces an asymmetry in the bistable potential equivalent to the addition of a constant driving term (Fig.~\ref{fig4}d). As $\delta$ is increased, we find that solitons and antisolitons now move at different terminal velocities in accordance with Eq.~\eqref{eq:NRDC_steadystate_velocity} and the Frenkel-Kontorova model (Fig.~\ref{fig4}e).

With differing velocities, solitons and antisolitons can now meet and collide (Fig.~\ref{fig4}f), contrary to the case of purely non-reciprocal driving, where solitons and antisolitons move at the exact same velocity.  In the presence of damping, such collisions have been shown to annihilate~\cite{Kosevich_Kivshar_paperInRussian,KivsharRevModPhys,Krasnov_PRB1998}, unlike their integrable counterparts ~\cite{dauxois2006physics}. Likewise, in our case, collisions result in annihilation of both excitations as a result of damping (see Extended Data Fig.~\ref{fig:SGcollision}), a phenomenon that one can exploit for various waveguiding applications.

We demonstrate this functionality by connecting a chain with a positively biased potential  of $+\delta$ a chain with a negative bias $-\delta$. When solitons and antisolitons are excited from the edge at small time intervals, the excitation with a higher velocity will catch up and annihilate before the interface between the two subsystems is reached. However, when the time interval is large enough, solitons and antisolitons do not catch up to each other, and arrive at the receiving end of the chain at the same intervals (Fig. \ref{fig4}g).

\begin{figure}[t!]
    \includegraphics[width=1\columnwidth,clip,trim=0cm 0cm 0cm 0cm]{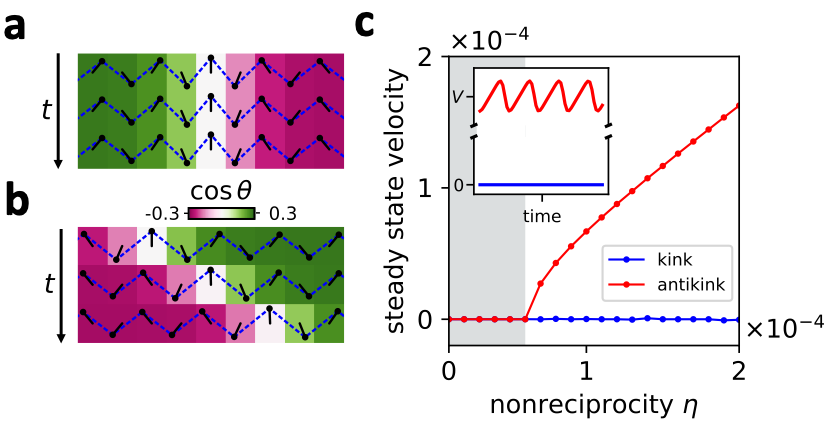}
 \caption{\textbf{Non-reciprocal solitons in the Kane-Lubensky chain}
(a) Soliton in the Kane-Lubensky chain remaining still in the presence of non-reciprocity for $\eta=1\cdot10^{-4}$.
(b) Antisoliton in the Kane-Lubensky chain being driven by non-reciprocity for $\eta=1\cdot10^{-4}$.
 (c) Steady state velocity of the soliton (blue) and antisoliton (red) as a function of non-reciprocity $\eta$. (inset) Total elastic energy $V$ vs. time for $\eta=1\cdot10^{-4}$. See Methods for details.} 
 \label{fig4new}
\end{figure}

Finally, we generalize our findings to another setting: the Kane-Lubensky chain, which is known to host $\phi^4$ solitons with zero energy while antisolitons have finite energy as a result of the half-breaking of the supersymmetry between these modes ~\cite{Chen2014,upadhyaya2020nuts}.
We extend this model to a non-reciprocal setting~\cite{Ghatak_PNAS2020} and focus on the overdamped regime (See Methods for details). Interestingly, we see that in the presence of non-reciprocity, solitons stay still (Fig.~\ref{fig4new}a) whereas antisolitons move (Fig.~\ref{fig4new}b). This asymmetry is due to the fact that solitons do not stretch springs while antisolitons do. The elastic energy of the antisolitons hence is finite and displays small oscillations as the antisolitons travel (Fig. ~\ref{fig4new}c-inset). These oscillations are due to the existence of a minute Peierls-Nabarro barrier that the antisoliton can overcome when driven by a small amount of non-reciprocity (Fig.~\ref{fig4new}c).
In conclusion, besides the broken symmetry between the solitons and antisolitons, the Kane-Lubensky chain reveals the same mechanism as in the case considered earlier, whereby non-reciprocal antisolitons sustain their own driving by imposing a local gradient of strain.


Summarizing, we have investigated how non-reciprocity and bistability can combine to stabilize excitations in an active mechanical metamaterial. 
This allows us to predict, control and manipulate the dynamic behavior of non-reciprocal topological solitons. It is an open question how the incommensurate phase of the Frenkel-Kontorova model and more generally geometric frustration~\cite{Xiaofei_Nature2023} and non-topological solitons such as breathers are affected by non-reciprocal driving. 
An interesting question is whether our findings have any bearing on defect dynamics in odd materials such as those reported in suspensions made of rotating particles, which interact non-reciprocally by virtue of hydrodynamic interactions~\cite{TanNature,Bililign2021Motile}. In the context of soft robotics, non-reciprocal topological solitons could provide exciting new avenues for autonomous and adaptable locomotion~\cite{Brandenbourger_arXiv2021}. More broadly beyond soft matter, it would be fascinating to realize non-reciprocal topological solitons in superconducting circuits~\cite{Baumgartner_NatNano2022}, quantum gases~\cite{Pucher_NatPhot} 
and optical microcavities~\cite{DelPino_Nature2022,Wanjura_NatPhys2023}.

\textit{Acknowledgments.} --- We thank Ronald Hassing and Kasper van Nieuwland for technical support and Freek van Gorp, Jean-S\'ebastien Caux, Jasper van Wezel, Anton Souslov, Jack Binysh and Vincenzo Vitelli for insightful discussions. We also thank Sibilla Bouché, Qianhang Cai and Jasper Lankhorst for gathering preliminary data in the context of the MSc course ``Project Academic Skills for Research'' they followed at the University of Amsterdam. We acknowledge funding from the European Research Council under grant agreement 852587 and from the Netherlands Organisation for Scientific Research under grant agreement VI.Vidi.213.131.3. 

\textit{Author contribution statement.} ---
C.C. and J.V. conceptualised and guided the project. J.V. and X.G. designed the samples and experiments. J.V. carried out the experiments. J.V., X.G. and C.C. carried out the numerical simulations. O.G., C.V.M. and A.S. performed the theoretical study. All authors contributed extensively to the interpretation of the data and the production of the manuscript. J.V. and C.C. wrote the Main Text. J.V. created the figures and videos. J.V., O.G., A.S. and C.V.M. wrote the Methods and the Supplementary Information.

\textit{Codes and data availability statement.} ---
All the codes and data supporting this study are available on the public repository \url{https://uva-hva.gitlab.host/published-projects/non-reciprocal-topological-solitons}.

\bibliography{biblio.bib}

\clearpage

\section{Methods}

\subsection{Experimental Methods}
Our active mechanical waveguide shown in Fig.~\ref{fig1}c consists of 50 3D printed rotating arms (with moment of inertia $I = 6.2 \pm 1.0 \cdot10^{-6} \mathrm{kg \ m^2} $) that are elastically coupled by rubber bands and positioned with a lattice spacing $a=6 \ \mathrm{cm}$. The rotating arms are coupled mechanically to a DC torque motor equipped with an angular decoder and a microcontroller that communicates with neighboring units to produce a external torque according to $\tau^a= \kappa^a ( \theta_{i-1} -  \theta_{i+1})$, identical to the experimental setup of ref. \cite{brandenbourger2019non}. 
To probe the response shown in Fig.~\ref{fig1}b, the system is excited at the edge by a short pulse of torque generated by the DC motor.
The bistable potential shown in the inset of Fig.~\ref{fig2} was constructed by attaching neodymium magnets to the tips of the oscillator arms and periodically spaced on an external substrate at distance of $x \ \mathrm{cm}$ from the rotor centre such that the potential minima are separated by an angle $\theta_d = 1 \ \mathrm{rad}$ (see Extended Data Fig. \ref{figS2}a). Although our metamaterial only supports two such minima, the results of the main text extend straightforwardly to the higher topological charge excitations that could potentially be generated and leveraged for more complex waveguiding in a multistable metamaterial, e.g. by using more intricate magnets layouts~\cite{Bifurcation_PNAS2023}. Extended data Fig.~\ref{fig:Multikink} shows simulations proving that solitons with larger topological charges exhibit the same robustness as the ones we investigate experimentally. The travelling solitons shown in Fig.~\ref{fig1}de were generated by initializing the chain with all sites sitting in the same mininum with the exception of the edge oscillator before turning on the non-reciprocal term.   
The Morse code message of Fig.~\ref{fig2} was generated by manually switching the oscillator arm at the edge from one stable state to the other at short (1 second) and long (3 second) intervals.

\subsubsection{Calibration of model parameters}

We model the active oscillator chain with a Frenkel-Kontorova model containing inertial, elastic, non-reciprocal, viscous, potential terms and constant force terms :
\begin{equation}
\begin{split}
I \pdv[2]{\theta_i}{\tau} \!= & \kappa (\theta_{i-1}\!+\!\theta_{i+1}\!-\!2\theta_{i})\!-\!\kappa^a (\theta_{i+1}\! -\! \theta_{i-1}) \!\\& -\! \gamma \pdv{\theta_i}{\tau} \!+\! B \sin(2\pi\frac{\theta_i}{\theta_d} )+E 
\end{split}
\label{eq:NRFKdim}
\end{equation}
By employing the following substitutions, we find the non-dimensional form of Eq.~\eqref{eq:NRFK}:
\begin{equation}
\begin{aligned}
    \phi_i &= 2\pi\frac{\theta_i }{\theta_d} +\pi \\
    t &= \sqrt{\frac{\kappa}{I}} \tau \\  
    \eta &= \frac{2\kappa^a}{ \kappa \sqrt{D}}\\
    \Gamma &= \frac{\gamma}{\sqrt{I \kappa D}} \\
    D &= \frac{2\pi B}{\kappa\theta_d} \\
    f &= \frac{2\pi E}{\kappa\theta_d} 
    \label{parameters}
\end{aligned}
\end{equation}

The elastic coupling $\kappa = 4.2\pm1.0 \cdot10^{-3} \mathrm{\ Nm \ rad^{-1}}$ (see Extended Data Fig. \ref{figS2}b) and the magnetic potential amplitude $B = 5.1\pm1.0\cdot10^{-4}  \mathrm{\ Nm \ rad^{-1}}$ (see Extended Data Fig. \ref{figS2}a) were calibrated by measuring the torques versus angle deviation on an Instron torsion testing machine. The relation between the magnet offset $\delta$ and the equivalent external force $E$ was found in the same way (see Extended Data Fig. \ref{figS2}de).
The viscous dissipation was found to be $\gamma = 2.0\pm0.5\cdot 10^{-4}\mathrm{\ Nm \ s \ rad}$ by fitting the oscillation amplitude decay after an initial perturbation (see Extended Data Fig. \ref{figS2}c). 
Shear bending forces in the elastic neighbor coupling were measured to be an order of magnitude smaller than the stretching forces in an earlier study \cite{brandenbourger2019non} and were thus neglected.
The speed of sound can be estimated through the lattice space $a$ as $c=a\sqrt{\kappa/I}$. Notice that after time and space rescaling \eqref{parameters} and taking the continuum limit (see below) the speed of sound is one: $c=1$. 
We remark here that Eq. \ref{eq:NRFK} has also been investigated \cite{Pinto-RamosPRL2021} in the context of the spontaneous formation of unstable $\pi$-solitons, but no experimental realizations have been investigated to our knowledge.
\subsection{Numerical Methods}

\subsubsection{Non-reciprocal Frenkel-Kontorova and sine-Gordon equations}

To verify the validity of the microscopic model, we found steady state (anti)soliton velocities by integrating Eq.~\eqref{eq:NRFK} with a velocity Verlet routine for a chain of length $N=512$ and using the parameter values and initial conditions as described above. For the ensuing dynamics, the sum of absolute angle deformation was used as a measure to differentiate diverging and dying out solutions from travelling soliton solutions. To find the soliton position, the field was then fitted to the continuum soliton solution given by $\phi = \pm 4 \arctan \exp \frac{x-vt}{\sqrt(1-v^2)}$ at each timestep, from which the steady state velocity as shown in \ref{fig2}ab was extracted by a linear fit.
In the continuum, the predicted steady state velocity given by Eq.~\eqref{eq:NR_dynamical_equation} were verified by integrating  Eq.~\eqref{eq:OddSineGordon} with the PyPDE package~\cite{py-pde} using the soliton solutions to the odd sine-Gordon equation as an initial condition on a grid of length $L=50$, spatial discretization $N=512$ and timestep $dt = 10^{-4}$.

\subsubsection{Non-reciprocal Kane-Lubensky chain}
Consider the Kane-Lubensky chain depicted in Extended data Fig.~\ref{fig:KL} and discussed in Fig.~\ref{fig4new} of the Main Text. This chain was first introduced in the context of topological insulators~\cite{KaneLubensky} and subsequently investigated in the nonlinear regime~\cite{Chen2014,Kinkantikink_KL,upadhyaya2020nuts}, where it was demonstrated to host solitons and antisolitons. Crucially solitons and antisolitons don't have the same energy: the existence of solitons does not require stretching any bonds while the antisolitons do. Such discrepancy has been proved to be associated to a 
half-breaking of the 
supersymmetry between the corresponding field equations, which are a supersymmetric version of the $\phi^4$ model~\cite{upadhyaya2020nuts}. The Kane-Lubensky chain has also been investigated in the non-reciprocal linear regime~\cite{Ghatak_PNAS2020} where it exhibits a non-Hermitian bulk-edge correspondence associated to the non-Hermitian skin effect. Here, we consider simultaneously the nonlinear and non-reciprocal regime of the Kane-Lubensky chain and ask how non-reciprocity drives solitons and antisolitons. With the parametrisation introduced in Extended Data Fig.~\ref{fig:KL}, the position of rotor $n$ is given by $\mathbf r_n=(\cos\theta_n,\sin\theta_n)$ and therefore since the rotors are staggered, the length of the spring connecting rotor $n$ to rotor $n+1$ is
\begin{equation}
\ell_n=\sqrt{\left(p-r c_n+r c_{n+1}\right)^2+\left(r s_n-r s_{n+1}\right)^2},
\end{equation}
where $c_n=\cos\theta_n$ and $s_n=\sin\theta_n$. Straining the springs induces then the elastic energy $V=(k/2)\sum_{n=1}^{N-1} (\ell_n-\ell_0)^2$, where $N$ is the number of rotors making up the chain, $\ell_0$ the rest length of the springs and $k$ the spring constant, which we fix to $k=1$ without loss of generality. In the reciprocal case, the torque on each rotor $n$ is given by $\tau_n^{\text{elastic}}=-\partial V/\partial\theta_n=\tau_n^L +\tau_n^R$, where $\tau_n^R=-(\ell_n-\ell_0) \partial\ell_n/\partial\theta_n$ is the torque exerted by the right adjacent spring and $\tau_n^L=-(\ell_{n-1}-\ell_0) \partial\ell_{n-1}/\partial\theta_n$ is the torque exerted by the left adjacent spring.
Here, we introduce non-reciprocity by adding an active term that introduces an asymmetry between these two torques 
$\tau_n^{\text{active}}=\eta(\tau_n^L-\tau_n^R$), where $\eta$ is the strength of the non-reciprocity. In the linear limit, such active forces precisely match those considered in~\cite{Ghatak_PNAS2020}. We then solve numerically the overdamped dynamics of such a chain given by the equations of motion
\begin{equation}
    \begin{split}\frac{d\theta_n}{dt}= 
    &
    -(1-\eta)(\ell_n-\ell_0) \frac{\partial\ell_n}{\partial\theta_n}\\
    &-(1+\eta)(\ell_{n-1}-\ell_0) \frac{\partial\ell_{n-1}}{\partial\theta_n}.
    \\
    \end{split}
\end{equation}
We consider two cases: (i) that of a soliton  (ii) and that of an antisoliton initially in the middle of the chain. We use the NDSolve solver of Mathematica and choose the following set of parameters $N=99$, $p=1$, $r=0.5$ and $\ell_0=\sqrt{p^2+4 r^2 \sin ^2\theta_0}$, where $\theta_0=\pi/2 +0.7$ for the soliton and $\theta_0=\pi/2 -0.7$ for the antisoliton. To prepare initial conditions, we first initialize the left half of the chain with $\theta_n=(\pi-(-1)^n\theta_0)$, the middle rotor with $\theta_{(N-1)/2}=\pi/2$ and the right half of the chain with $\theta_n=-(-1)^n\theta_0$. We then let the system relax under overdamped dynamics and use the relaxed configuration as an initial condition. The results are displayed in Fig.~\ref{fig4new} of the Main Text. Interestingly, since solitons don't stretch any spring, non-reciprocal driving is not able to drive them and they remain still. Only antisolitons are driven by non-reciprocity. Also, since the Peierls-Nabbaro barrier is very small~\cite{Kinkantikink_KL}, the threshold reciprocity to accelerate the antisoliton is also very small.

\subsection{Theoretical Methods}

\subsubsection{Continuum limit}

The continuum limit of Eq.~\eqref{eq:NRFK} is found by letting $\phi_i$ become a continuous function $\phi(x)$ of space $x \in [0,Na]$, where $N$ is the number of units. Approximating finite differences by a Taylor expansion according to
${    \phi_{i+1}-\phi_i \approx   a \phi_x + a^2\phi_{xx}/2}$
and substituting terms in the discrete model of Eq.~\eqref{eq:NRFK} then leads to Eq.~\eqref{eq:OddSineGordon} under rescaling of the spatial variable $x \rightarrow \frac{a}{\sqrt{D}}x$ and time $t\rightarrow \frac{t}{\sqrt{D}}$. In these units the speed of sound is $c=1$.

We note here that earlier work treats a special case of Eq.~\eqref{eq:OddSineGordon} where the model parameters $\eta$ and $\Gamma$ are spatially varying functions and the systems described are not translationally invariant. Consequently, the (anti)soliton kinetic energy is not constant but gradually vanishes as it travels along the stiffness grading~\cite{Arrieta2018} or the potential grading~\cite{KatiaNatComm}. This decrease in velocity precludes the possibility of efficient waveguiding when these systems are scaled up. In addition, we emphasize here that systems with stiffness or potential grading are inherently constrained to a finite size since practical limitations on material properties and manufacturing forbid gradings from becoming arbitrarily small.

\subsubsection{Stability analysis}

That solitons are stable does not guarantee that all solutions to Eq.~\eqref{eq:OddSineGordon} are (Extended Data Fig.~\ref{fig:stability}a). 
The threshold of stability of radiative modes can be predicted by analyzing the stability of perturbations around the soliton profile travelling at the speed of sound (defined in the unperturbed linear system), in the limit of $v \to 1$. 
The dispersion relation for such solutions yields the following complex frequencies (see Methods section `Perturbative excitations' below for details):
\begin{equation}
    \omega_\pm =- \frac{i \Gamma}{2} \pm \sqrt{1+k^2 - (\Gamma/2)^2 + i k \eta}.
    \label{dispersion_above_threshold}
\end{equation}
The growth rates of perturbations given by $\Im(\omega_\pm)$ become positive for $\eta\ge\Gamma$ starting with the highest wavenumbers $k$ (Extended Data Fig.~\ref{fig:stability}b). 
Numerical integration of Eq. (\ref{eq:OddSineGordon}) in the supersonic limit confirms the generation of exponentially amplified high wavenumber modes (Extended Data Fig.~\ref{fig:stability}a). These unstable modes indicate that non-reciprocal topological solitons driven beyond the speed of sound can no longer dissipate sufficiently, causing excess energy to build up exponentially---reminiscent of the sonic boom experienced by an object breaking the sound barrier.

Since the speed of sound in a material is inversely proportional to its mass density, solitons are expected to always be stable in the overdamped limit, as we show by repeating the above analysis (see perturbative excitations section below).
Since we are concerned here with the small amplitude limit and only describe (anti)solitons of topological charge $\pm 1$, a non-reciprocal $\phi^4$ model should also suffice to capture soliton dynamics. In the S.I., we treat this model perturbatively and  show that the main results hold.

\subsubsection{Inverse scattering transform}

In this chapter, we briefly describe the derivation of Eq. \eqref{eq:NR_dynamical_equation}. 
To be more general we also include a constant driving term, so Eq. \eqref{eq:OddSineGordon} takes the following form
\begin{equation}
	\pdv[2]{\phi}{t}-\pdv[2]{\phi}{x}+\sin \phi = -\eta\pdv{\phi}{x} - \Gamma\pdv{\phi}{t}+f\equiv R[\phi].
	\label{eq:OddSineGordonDC}
\end{equation}
In case $R[\phi]=0$, the equation turns out to be integrable and its solutions can be found by the inverse scattering procedure \cite{Faddeev_1987}.
Namely, one has to first find a scattering matrix for the linear problem whose potential depends on the field configuration $\phi$ and its derivatives in the initial moment of time
\begin{equation}\label{jost2}
	\frac{d T_{\pm}(x,\lambda)}{dx} = U T_{\pm}(x,\lambda)
\end{equation}
where the $2\times 2$ matrix $U$ depends on the spectral parameter $\lambda$
\begin{equation}
	U =\frac{\partial_t \phi \sigma_3 }{4i} + \frac{\lambda+\lambda^{-1}}{4i} \sigma_1\sin \frac{\phi }{2}
	+ \frac{\lambda-\lambda^{-1}}{4i} \sigma_2\cos \frac{\phi }{2},
\end{equation}
and solutions $T_\pm$ are specified by their behaviour at ${x\to\pm \infty}$. They are called the Jost solutions and differ from each other by multiplication on the constant scattering or transfer matrix $T(\lambda)$ 
\begin{equation}\label{ttSG}
	T_- (x,\lambda) = T_+(x,\lambda) T(\lambda),\quad T(\lambda) = \left(	\begin{array}{cc}
		a(\lambda) & - \bar{b}(\lambda) \\
		b(\lambda) & \bar{a}(\lambda)
	\end{array}
	\right).
\end{equation}
For example, a soliton profile parametrized by a real positive parameter $\kappa$ has a form
\begin{equation}\label{prof}
	\phi(x,t) = -4 \arctan (e^{x(\kappa+1/\kappa)/2}/\gamma(t)),
\end{equation}
where evolution of $\gamma(t)$ is given by 
\begin{equation}\label{gammat0}
    \gamma(t) = e^{-t(\kappa-1/\kappa)/2} \gamma_0,
\end{equation}
gives the following Jost solutions at $t=0$
\begin{equation}\label{JostSG}
	T_+ = \frac{\mathcal{E}}{\sqrt{1+e^{2\xi}}}\left(\begin{array}{cc}
		\frac{\lambda+ i \kappa}{\lambda - i \kappa}  & -e^{\xi} \\
		e^{\xi}& \frac{\lambda- i \kappa}{\lambda + i \kappa} 
	\end{array}\right)e^{-i\sigma_3 x \frac{\lambda^2-1}{4\lambda}}
\end{equation}
\begin{equation}\label{JostSG1}
	T_- = \frac{\mathcal{E}}{\sqrt{1+e^{2\xi}}}\left(\begin{array}{cc}
		1 & -\frac{\lambda+ i \kappa}{\lambda - i \kappa}e^{\xi} \\
		\frac{\lambda- i \kappa}{\lambda + i \kappa}e^{\xi}&  1
	\end{array}\right)e^{-i\sigma_3 x\frac{\lambda^2-1}{4\lambda}}.
\end{equation}
Here the constant matrix $ \mathcal{E}$ and parameter $\kappa$ are given by
\begin{equation}
	\mathcal{E} = \frac{1}{\sqrt{2}} \left(
	\begin{array}{cc}
		1 & i \\
		i & 1
	\end{array}
	\right),\quad v  = \frac{1-\kappa^2}{1+\kappa^2},
\end{equation}
and $e^{\xi}=-e^{x(1+\kappa^2)/(2\kappa)}/\gamma_0$. If $\gamma_0<0$ such a solution is called a soliton and if $\gamma_0>0$ - an antisoliton.  
In both cases the corresponding transfer matrix is diagonal 
\begin{equation}
	a(\lambda) = \frac{\lambda-i\kappa}{\lambda+i\kappa},\qquad b(\lambda)=0,
\end{equation}
The quantity $\gamma_0$ should be regarded as additional scattering data, defined in the general situation as a propotionality coefficient 
between the first column of $T_-$ and the second column of $T_+$ for the spectral parameter $\lambda_k$ that is a zero of the $a(\lambda)$ in the upper half plane. i.e. $a(\lambda_k)=0$, ${\rm Im}\lambda_k>0$
\begin{equation}
	T_-^{(1)}(x,\lambda_k) = \gamma_k T_{+}^{(2)}(x,\lambda_k),\qquad k = 0, 1, \dots ,n
\end{equation}
The dynamics of the scattering data is extremely simple 
\begin{equation}\label{dyn2SG}
	a(\lambda,t) = a(\lambda,0),\quad b(\lambda,t) = e^{it(\lambda^2+1)/(2\lambda)}b(\lambda,0), 
\end{equation}
\begin{equation}\label{dyn2SG1}
	\lambda_k(t) = \lambda_k(0),\quad \gamma_k(t) = e^{it(\lambda_k^2+1)/(2\lambda_k) } \gamma_k(0).
\end{equation}
After this evolution the time dependence of the profile can be recovered via the inverse scattering transformation \cite{Faddeev_1987}. 

For $R[\phi]\neq 0$ for one-soliton case we can use perturbation theory in the adiabatic approximation, which means that the form of the profile still reads as   Eq. \eqref{prof}, but the evolution \eqref{gammat0} is modified along with the other soliton's parameters. More precisely, one can demonstrate the following evolution of the transfer matrix
\begin{equation}\label{pert1}
	\frac{dT(\lambda)}{dt} - i\frac{\lambda^2+1}{4\lambda} [\sigma_z ,T(\lambda)] =\int\limits_{-\infty}^{\infty} \frac{dz}{4i} T_+^{-1}(z)\hat{R}[z]T_-(z).
\end{equation}
\begin{multline}\label{pert2}
	\frac{d\gamma}{dt}- \frac{1-\kappa^2}{2\kappa } \gamma  = \\  \frac{i}{\dot{a}(i\kappa)}\int\limits_{-\infty}^{\infty} \frac{dz}{4i}  \left[\dot{T}^{(1)}_-(z) -\gamma\dot{T}^{(2)}_+(z) \right]^{T}\sigma_2\hat{R}[z]T^{(1)}_-(z) 
\end{multline}
\begin{equation}\label{pert3}
	i\frac{d\kappa}{dt}  =  \frac{i}{\dot{a}(i\kappa)}\int\limits_{-\infty}^{\infty} \frac{dz}{4i}  \left[T_+^{(2)}(z)\right]^{T}\sigma_2\hat{R}[z]T^{(1)}_-(z) 
\end{equation}
Here $\hat{R}[z]=R[\phi(z)]\sigma_3$, dot means derivative over a spectral parameter $\lambda$, and right part of Eqs. \eqref{pert2} and \eqref{pert3} should be evaluated at $\lambda  = i \kappa$. Using (\ref{JostSG},\ref{JostSG1}) we obtain
\begin{equation}\label{kappa}
	\frac{d \kappa}{dt} =- \frac{\Gamma  \kappa  \left(\kappa ^2-1\right)}{\kappa ^2+1}-\eta  \kappa - \frac{\pi f {\rm sgn} (\gamma_0)}{2} \frac{\kappa^2}{1+\kappa^2}
\end{equation}
\begin{equation}
	\frac{d \gamma}{dt} = \frac{1-\kappa^2}{2\kappa } \gamma  - \frac{\gamma \log(\gamma^2)}{2} 
	\frac{1-\kappa^2}{\kappa(1+\kappa^2)} \frac{d\kappa}{dt} 
\end{equation}
Once $\gamma$ and $\kappa$ are found the profile can be recovered from Eq. \eqref{prof}. 
Notice that only appearance of the force $f$ makes a distinction between soliton and antisoliton. 
Let us focus on $\gamma_0>0$. 
And introduce new variables 
\begin{equation}
	\log \gamma= \frac{\kappa + 1/\kappa}{2} X_c (t),\qquad 	W(t) = \frac{2}{\kappa + 1/\kappa}.
\end{equation}
which leads to the following form of the profile
\begin{equation}\label{width2}
	\phi(x,t) = -4 \arctan\exp\left(
	\frac{x-X_c(t)}{W(t)}
	\right)
\end{equation}
with $\gamma_0$ included in $X_c(0)$. 
Dynamics for $X_c(t)$ allows us to define the velocity
\begin{equation}
	\frac{dX_c}{dt} = v=\frac{1-\kappa^2}{1+\kappa^2} 
\end{equation}
and 
\begin{equation}
	\frac{dv}{dt} = -\Gamma (v - v_\eta)(1-v^2) + \frac{\pi f}{4}(1-v^2)^{3/2}
\end{equation}
here $v_\eta =  \eta /\Gamma$. The critical points can be easily found from \eqref{kappa} 
\begin{equation}
	\eta+\Gamma  -(\eta -\Gamma)\kappa^2 - \frac{\pi f}{2}\kappa = 0.
\end{equation}
The answer for soliton will result in flipping the sign of the force. This way, we obtain the following velocities for soliton and antisoliton: 

For antisoliton: 
\begin{equation}
	v_{t} = \begin{cases}
		v_+,  & -\Gamma\le \eta \le \Gamma \\
		v_+ & -\sqrt{\Gamma^2 + \pi^2 f^2/16} \le \eta \le -\Gamma, \qquad v(0)>v_- \\
		-1 & -\sqrt{\Gamma^2 + \pi^2 f^2/16}\le \eta \le-\Gamma, \qquad v(0)<v_- \\
		-1, & \eta<-\sqrt{\Gamma^2 + \pi^2 f^2/16} \\
		+1, & \eta>\Gamma
	\end{cases}
\end{equation}
For soliton: 
\begin{equation}
	v_{t} = \begin{cases}
		v_-,  & -\Gamma\le \eta \le \Gamma \\
		v_- & \Gamma\le \eta \le \sqrt{\Gamma^2 + \pi^2 f^2/16}, \qquad v(0)<v_+ \\
		+1 & \Gamma\le \eta \le \sqrt{\Gamma^2 + \pi^2 f^2/16}, \qquad v(0)>v_+ \\
		+1, & \eta>\sqrt{\Gamma^2 + \pi^2 f^2/16} \\
		-1, & \eta<-\Gamma
	\end{cases}
\end{equation}

where 
\begin{equation}
	v_{\pm} = \frac{\pm \frac{\pi f}{4}\sqrt{\Gamma^2 - \eta^2 + \pi^2 f^2/16}+\eta \Gamma}{\Gamma^2 + \pi^2 f^2/16}.
\end{equation}
Notably in the absence of the force $f$ there is no difference in the finite velocity for the soliton or antisoliton
\begin{equation}
	v_+ = v_- = \eta /\Gamma.
\end{equation}

\subsubsection{Perturbative excitations}

Let us also discuss the role of perturbative excitations on top of the soliton-like profile $\phi_k$.
By shifting $\phi \to \phi_k + \phi$ in Eq. \eqref{eq:OddSineGordonDC} and keeping only linear terms in $\phi$ we obtain 
\begin{equation}
	\partial_t^2 \phi - \partial_x^2 \phi+ \eta \partial_x \phi +\Gamma \partial_t\phi  + \phi  = \phi V + G
\end{equation}
here the driving $G$ and the potential $V$ are local functions and do not play a role in the continuous spectrum, but might be responsible for the localized bound states modes which we extensively studied in \cite{Kivshar1998}. 
So for continuous spectrum, we study the following equation 
\begin{equation}
	\partial_t^2 \phi - \partial_x^2 \phi+ \eta \partial_x \phi +\Gamma \partial_t\phi  + \phi  = 0.
\end{equation}
The plane wave ansatz 
\begin{equation}
	\phi(x,t) = e^{ikx-i\Omega t} 
\end{equation}
with real $k$, leads to the following equation for $\Omega$ 
\begin{equation}
	\Omega^2 +i \Omega \Gamma -i k \eta  -k^2-1=0,
\end{equation}
which gives the following frequencies
\begin{equation}
	\Omega_\pm =- \frac{i \Gamma}{2} \pm \sqrt{1+k^2 - (\Gamma/2)^2 + i k \eta}.
\end{equation}
The stability regions are defined by the condition ${\rm Re}(-i \Omega) = {\rm Im} \Omega<0$. 
Notice that as $k \to -k $, $ {\rm Im} \Omega_+ \to   {\rm Im} \Omega_- $. 
In Extended Data Fig. \ref{fig:stability}c, we plot the imaginary parts of $\Omega_\pm$ for various values of $\eta$ for $\Gamma=1$, showing that for $|\eta|>\Gamma$ there will be an instability region, namely for 
\begin{equation}
	|k|>\frac{\Gamma}{\sqrt{\eta^2 - \Gamma^2}} 
\end{equation}
either $\Omega_+$ or $\Omega_-$ will have a positive imaginary part.

In the overdamped regime, where $\partial_t^2 \phi$ can be neglected the dispersion simplifies to

\begin{equation}
	\Omega = k \frac{\eta}{\Gamma} -i\frac{1 +  k^2}{ \Gamma}  
\end{equation}

such that the imaginary part is always negative, meaning that solitons are always stable.

\setcounter{figure}{0}
\renewcommand{\figurename}{Extended Data FIG.}

\begin{figure*}[h!]
  \includegraphics[width=2\columnwidth,clip,trim=0cm 0cm 0cm 0cm]{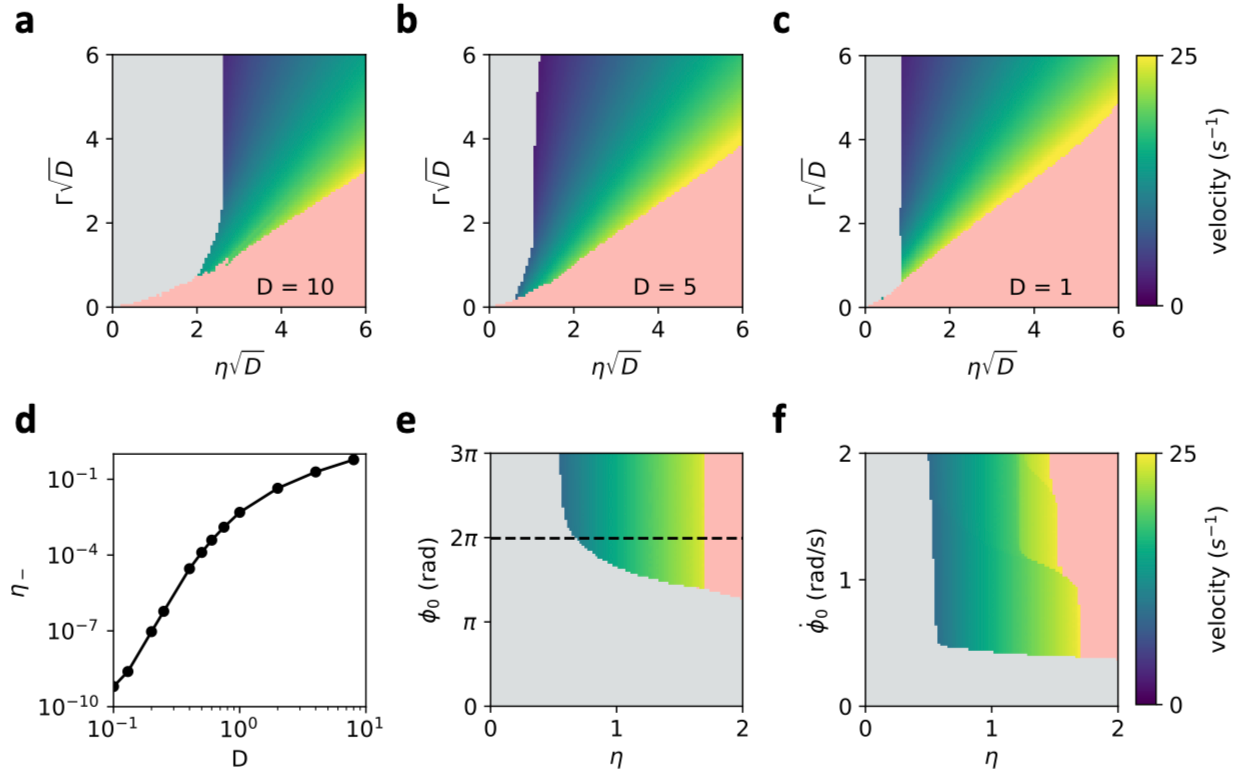}
 \caption{\textbf{Dependence of the Peierls-Nabarro barrier on the nondimensional amplitude $D$ and initial conditions in the Frenkel-Kontorova model} (\textbf{abc}) The Peierls-Nabarro barrier, regime of instability and (anti)soliton velocities as the continuum limit is approached as a function of the unnormalized non-reciprocity $\eta \sqrt{D}$ and damping $\Gamma \sqrt{D}$. 
 As the discreteness parameter $D$ becomes smaller, the line separating stable from unstable solutions approaches $\Gamma=\eta$ as predicted for the continuum. The initial condition used corresponds to the experimentally used soliton with single lattice spacing width. In addition, the Peierls-Nabarro barrier gradually decreases and (\textbf{d}) eventually goes to zero, provided that the initial soliton shape also becomes less discrete \cite{PEYRARD198488}. (\textbf{e})  When the activation amplitude $\phi_0$ of the experimental initial condition is changed, the Peierls-Nabarro barrier also changes but for large enough amplitudes, it becomes constant.
 (\textbf{f}) When instead of an initial activation angle, an edge oscillator is initialized with some radial velocity $\dot{\phi}_0$, the Peierls-Nabarro barrier remains constant.  }
 \label{fig:PN_dependence}
\end{figure*}
\begin{figure*}[h!] 
	\centering
	\includegraphics[width=1\linewidth]{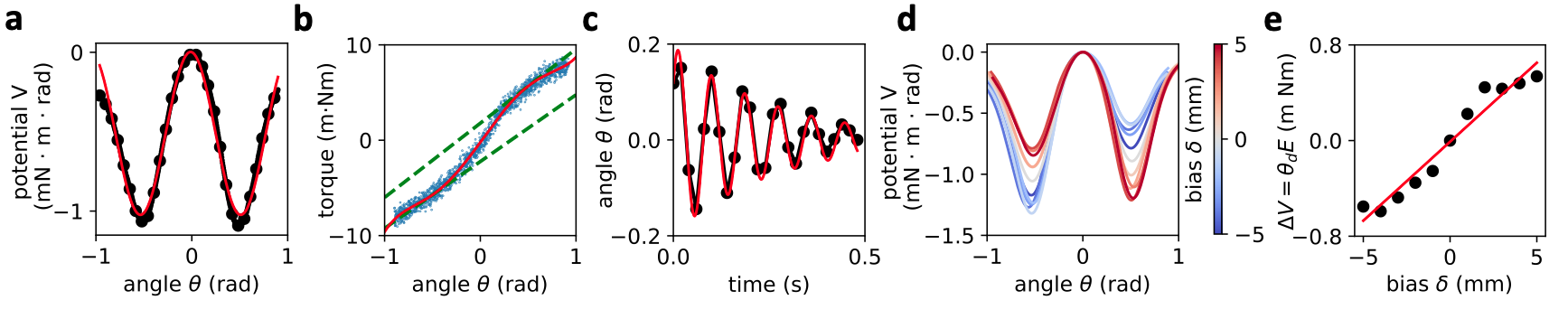} 
	\caption{\textbf{Calibration of experimental parameters} \textbf{(a)} The nonlinear potential generated by the periodically spaced magnets, as measured with an Instron torsion testing machine. Red line represents the sinusoidal fit used to calibrate the magnetic potential amplitude $B$.
\textbf{(b)} Instron measurement of the elastic forces experienced by a single oscillator connected to two neighboring oscillators. Red line shows the smoothed data and green dashed lines show linear fits around the two potential minima, denoting the elastic coupling strength $\kappa$.
 \textbf{(c)} Oscillation of a single oscillator elastically coupled to two neighbors, used to measure the viscous damping coefficient $\gamma$.
 \textbf{(d)} The biased potential for different amounts of bias $\delta$. 
 \textbf{(e)} The difference between the potential minima $\Delta V$ between the two uneven minima plotted versus the bias $\delta$. A linear fit establishes the relation between the bias and $\delta$ the effective external force $E$ it corresponds to. 
 }
 \label{figS2}
\end{figure*}

\begin{figure*}[h!] 
	\centering
	\includegraphics[width=1\linewidth]{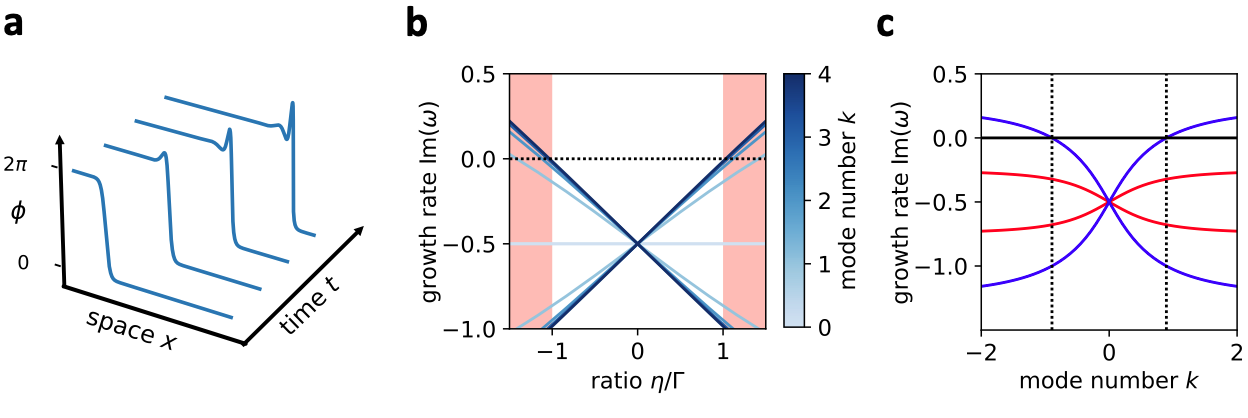}
	\caption{\textbf{Stability of the soliton} (\textbf{a}) Snapshots of a soliton in the unstable regime showing the destabilization of high wavenumber modes, found numerically for $\eta=1.1$ and $\Gamma=1$.
 (\textbf{b}) Growth rates $\Im (\omega)$ of perturbations around the soliton solution for various wavenumbers given by Eq. (\ref{dispersion_above_threshold}). The dotted line at $\Im(\omega)$ marks the transition between decaying and growing solutions, with high wavenumbers being the first to become unstable as the threshold of stability $\eta=\Gamma$ is crossed. (\textbf{c}) Dependence of ${\rm Im}\Omega_\pm$  on the wavenumber $k$ for $\Gamma=1$ and $\eta =0.5$ (red) and $\eta=1.5$ (blue). In the latter case, modes in the regions ${\rm Im}\Omega_+>0$ become unstable at $k=\pm \Gamma/\sqrt{\eta^2-\Gamma^2}$ given by the dashed lines.}
	\label{fig:stability}
\end{figure*}

\begin{figure*}[t!]
 \begin{center}
  \includegraphics[width=1\columnwidth,clip,trim=0cm 0cm 0cm 0cm]{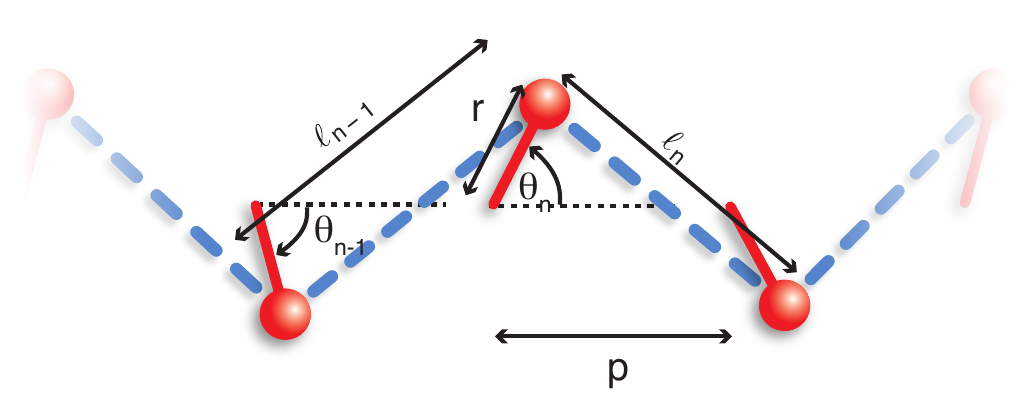}
 \end{center}
 \caption{{\bf{Kane-Lubensky chain.}}  
 Sketch of the Kane-Lubensky chain and its notation conventions.
}
 \label{fig:KL}
\end{figure*}

\begin{figure*}[t!]
 \begin{center}
  \includegraphics[width=1\columnwidth,clip,trim=0cm 0cm 0cm 0cm]{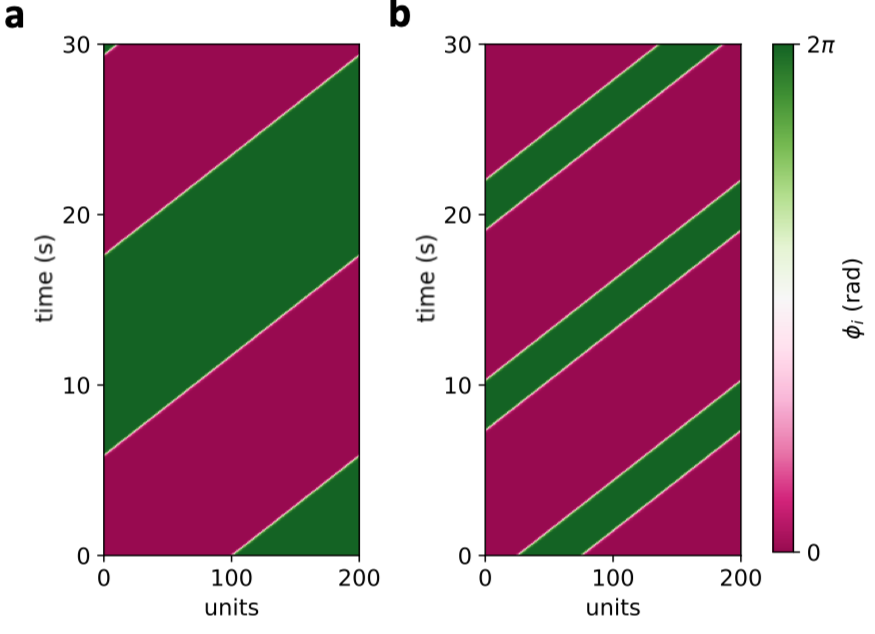}
 \end{center}
 \caption{\textbf{Insensitivity of non-reciprocal solitons to boundary conditions}. Although at a linear level, the non-Hermitian skin effect causes the energy spectrum to change radically upon changing boundary conditions, nonreciprocal solitons are insensitive to the boundary as their topological charge protects them from amplifying exponentially in space.   (\textbf{a}) Simulation of a single Frenkel-Kontorova soliton driven by non-reciprocity ($\eta=1.1$, $\Gamma = 1.3$, $D=1.2$ ) under antiperiodic boundary conditions.  (\textbf{b}) Simulation of a Frenkel-Kontorova soliton-antisoliton pair driven by non-reciprocity ($\eta=1.1$, $\Gamma = 1.3$, $D=1.2$ )  under periodic boundary conditions. Neither periodic, antiperiodic or the open boundary conditions used in the main text affect the stability and velocity of the (anti)soliton.
}
 \label{fig:PBC}
\end{figure*}

\begin{figure*}[t!]
 \begin{center}
  \includegraphics[width=0.5\columnwidth,clip,trim=0cm 0cm 0cm 0cm]{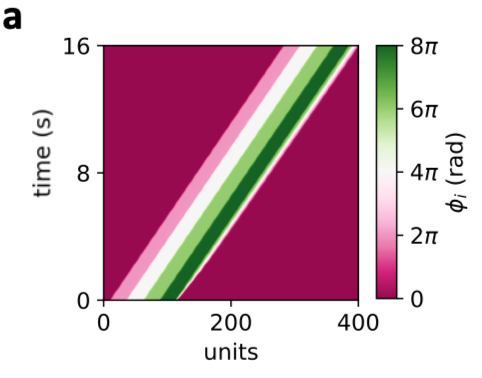}
 \end{center}
 \caption{\textbf{Solitons with higher topological charge} (\textbf{a}) Simulation of a staircase of Frenkel-Kontorova solitons under the influence of non-reciprocity ($\eta=1.1$, $\Gamma = 1.3$, $D=1.2$ ). As in the single soliton case, (anti)solitons with higher topological charge travel undisturbed at the same steady state velocity.  }
 \label{fig:Multikink}
\end{figure*}

\begin{figure*}[t!]
 \begin{center}
  \includegraphics[width=1\columnwidth,clip,trim=0cm 0cm 0cm 0cm]{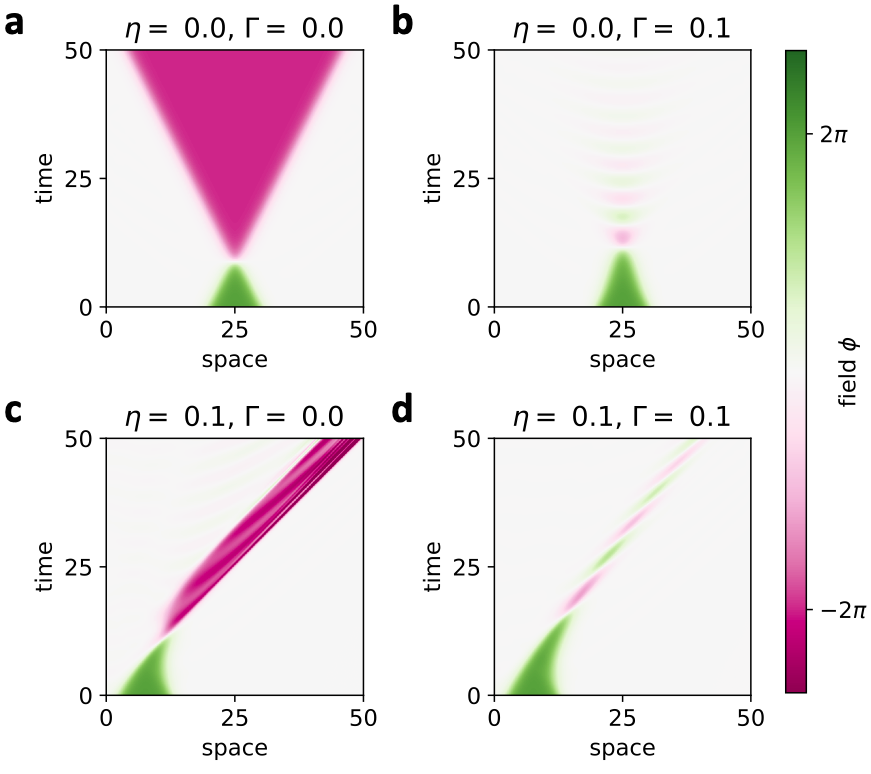}
 \end{center}
 \caption{\textbf{Effect of non-reciprocal driving and damping on the collision of sine-Gordon solitons} (\textbf{a}) In the absence of both driving and damping, solitons and antisolitons pass through each other without interacting. (\textbf{b}) For nonzero damping, soliton and antisoliton annihilate and the resulting non-topological solution dissipates away. (\textbf{c}) With only non-reciprocity turned on, both excitations still pass through each other unhindered but are also rendered unstable. (\textbf{d}) Dissipation and non-reciprocity can balance, giving rise to non-reciprocal breather solutions.
 }
 \label{fig:SGcollision}
\end{figure*}

\end{document}